# TRANSVERSE POLARIZATION FOR ENERGY CALIBRATION AT THE Z PEAK

M. Koratzinos, University of Geneva and CERN, Geneva, Switzerland

*Abstract*

In this paper we deal with aspects of transverse polarization for the purpose of energy calibration of proposed circular colliders like the FCC-ee and the CEPC. The main issues of such a measurement will be discussed. The possibility of using this method to accurately determine the energy at the WW threshold as well as the Z peak will be addressed. The use of wigglers for reducing long polarization times will be discussed and a possible strategy will be presented for minimising the energy uncertainty error in these large machines.

## INTRODUCTION

Accurate energy determination is a fundamental ingredient of precise electroweak measurements. In the case of LEP1 the centre of mass energy at and around the Z peak was known with an accuracy of around $2\times10^{-5}$. The exact contribution of the energy error to the mass and the width of the Z are presented in [1].

The proposed circular colliders FCC-ee [2] and CEPC [3] are capable of delivering statistics a factor $\sim10^5$ larger than LEP at the Z and WW energies, therefore there is a need not only to achieve similar performance as far as energy determination is concerned, but to do significantly better.

The only method that can provide the accuracy needed is the so-called resonant depolarization technique, each measurement of which has an instantaneous accuracy of $O(10^{-6})$.

The resonant depolarization technique [4] is based on the fact that the spin precession frequency of an electron in a storage ring is proportional to its energy, $E$. More precisely the spin tune $\nu$ will precess $a\gamma$ times for one revolution in the storage ring, where $a$ is the anomalous magnetic moment and $\gamma$ the Lorenz factor of the electron

$$\nu = \alpha\gamma = \frac{aE}{mc^2} = \frac{E[MeV]}{440.6486(1)[MeV]} \quad (1)$$

The average of all spin vectors in a bunch is defined as the polarization vector $\vec{P}$. Therefore the average energy of a bunch can be computed by selectively depolarizing a bunch of electrons or positrons which have been polarized to an adequate level and measuring the frequency at which this depolarization occurs. A polarimeter measures the change of polarization level. The accuracy with which the instantaneous average energy of the bunch is computed using this method is O(100KeV) – a value much smaller than the beam energy spread of the storage rings considered here.

## TRANSVERSE POLARIZATION

Electron and positron beams in a storage ring naturally polarize due to the Sokolov-Ternov effect [5]. For the purposes of energy calibration important figures of merit are the asymptotic value of polarization that can be reached and the time constant of polarization build-up.

### Asymptotic polarization value

The maximum achievable polarization value is given by the theory as

$$P_{max} = \frac{8}{5\sqrt{3}} \cong 0.924 \quad (1)$$

however machine imperfections usually limit this number to lower levels. There can be numerous depolarizing effects in a storage ring.

### Polarization time constant

For a beam with zero polarization the time dependence for build up to equilibrium is

$$P(t) = P_{max}[1 - \exp(-t/\tau_{pol})] \quad (2)$$

Where the built up rate is (in natural units)

$$\tau_{pol}^{-1}[s^{-1}] \approx \frac{2\pi}{99}\frac{E[GeV]^5}{C[m]\rho[m]^2} \quad (3)$$

Where C is the circumference of the storage ring and $\rho$ its bending radius. Therefore polarization times increase dramatically with the machine circumference and decrease with energy. The use of wigglers [6] can decrease this time if needed, at the expense of increasing the energy spread and the synchrotron radiation (SR) budget of the machine.

Polarization times for relevant machines and energies can be seen in Table 1.

**Table 1:** Polarization times without the help of wigglers

| Storage ring | Circumference (kms) | E (GeV) | $\tau_{pol}$ (hours) |
|---|---|---|---|
| LEP | 27 | 45 | 5.8 |
| FCC-ee | 100 | 45 | 290 |
| FCC-ee | 100 | 80 | 16 |
| CEPC | 55 | 45 | 48 |

## POLARIZATION AND ENERGY SPREAD

One important limitation on achievable polarization levels comes from the energy spread of the beam: Off momentum particles reaching an integer spin resonance depolarize quickly. As seen from eqn. (1), these spin resonances sit 440MeV apart. Energy spread scales approximately like

$$\sigma_E \propto \frac{E^2}{\sqrt{\rho}} \quad (4)$$

The maximum energy at which useful levels of polarization can be measured cannot easily be calculated. However, we can extrapolate from the measurements done at LEP [7] where the maximum energy where polarization was observed was 60.6GeV (at a level of around 8%). Using eqn. (4) this extrapolation to storage rings with different diameters can be seen in Table 2.

**Table 2:** extrapolation of LEP data to other machines regarding the maximum energy below which polarization levels will be adequate for resonant depolarization measurements

| Storage ring | C(kms) | Maximum energy with polarization (GeV) |
|---|---|---|
| LEP | 27 | 61 |
| CEPC | 55 | 72 |
| FCC-ee | 100 | 84 |

As can be seen from the above table, polarization at the W pair threshold (80GeV) at FCC-ee seems possible. This is in contrast of what was achieved at LEP and another input to the physics case of this unique machine.

There are also LEP measurements (figure 8 in [7]) where the maximum energy spread compatible with reasonable polarization levels has been measured - wigglers were used to change the energy spread. An energy spread larger than about 52MeV leads to a significant drop of polarization levels. In the absence of detailed simulation work we shall use the above figure as the maximum permissible energy spread compatible with polarization.

## RESONANT DEPOLARIZATION

The way the resonant depolarization measurement is performed is the following: Only one bunch is targeted at a time. Since the colliding rate is much larger than the polarization rate, for polarization to build up this bunch needs to be a non-colliding bunch. It should be stated here that operation with colliding and non-colliding bunches might be a challenge due to the different tune shifts of the two species of bunches involved. The measurement proper consists of measuring the spin precession frequency by introducing a resonance in a 'trial and error' fashion. If no depolarization is observed (failure) the frequency used is not the correct depolarizing frequency. The bunch remains polarized. If the bunch depolarises (success) the frequency corresponds to the exact mean energy of the bunch at that moment. To observe the polarization change, polarization levels of 5-10% are needed – the better the polarimeter, the lower the values of polarization necessary for a successful measurement.

## THE ENERGY MODEL

The beam energy of large storage rings continuously changes due to internal and extraneous causes. This evolution can be modelled but energy changes are many orders of magnitude larger than the instantaneous accuracy of a depolarization measurement. For example, small changes in the diameter of the ring due to elastic deformations of the earth's crust (due to, for instance, tidal forces) can have a big effect on the energy of the electrons and positrons. This is due to the small momentum compaction factor $\alpha_c$ which relates changes in energy to changes in the orbit length of a storage ring:

$$\frac{\Delta E}{E} = -\frac{1}{\alpha_c}\frac{\Delta L}{L} \quad (5)$$

Where $L$ is the orbit length. Momentum compaction factors vary from $2 \cdot 10^{-4}$ for LEP to $5 \cdot 10^{-6}$ at FCC-ee. So a 1mm orbit length change at FCC-ee (a change of $10^{-8}$) leads to a large 90MeV change in energy, to be contrasted with the O(100keV) accuracy of the depolarization method. Table 3 shows changes in energy for a 1mm circumference change (typical for tide-induced changes at LEP) for the three storage rings discussed here.

**Table 3:** change in energy of a 45GeV beam for a circumference change of 1 mm in the three storage rings discussed here

| Storage ring | Circumference (kms) | $\alpha_c$ | $\Delta E$ (MeV) |
|---|---|---|---|
| LEP | 27 | $2 \cdot 10^{-4}$ | 8 |
| CEPC | 55 | $4 \cdot 10^{-5}$ | 20 |
| FCC-ee | 100 | $5 \cdot 10^{-6}$ | 90 |

The time constant for elastic deformation changes varies between hours (for tides) to months (for rainfall variations). Other effects that contribute to change of the energy include temperature changes (time constant of a few hours), parasitic currents (time constant of some minutes) and many other effects discussed in some detail in [4].

Moreover, the RF configuration can give rise to different energies for electrons and positrons, therefore both species should be measured with the resonant depolarization technique, something that was not done at LEP. This necessitates the use of two polarimeters for both species.

There are also corrections to be applied in deriving the centre of mass energy per experiment from the mean energy of the electrons and positrons measured with the resonant depolarization method.

Therefore for ultimate precision we need to

- Measure the energy using the resonant depolarization every few minutes
- Measure independently electrons and positrons
- Measure continuously from the beginning of physics to the end of physics

## WIGGLERS

The natural polarization time for large rings is very long as seen from Table 1. As we only need polarization levels of 5-10% to perform a polarization measurement, we can divide the numbers in the table by 10 to 20. But this is still too long compared to the mean time between failures which cannot be assumed to be more than a few hours or a day at most. A way to reduce polarization time is the use of wigglers [6]. Wigglers are dipole magnets with two parts: a low field region and a high field region so that the integral field seen by the electrons is zero. However they help polarization as polarization time scales with the square of the field and polarization levels are not affected provided that the wiggler asymmetry (the ratio of lengths of the positive and negative field magnets) is larger than ~5.

Wigglers have, however, two undesired effects: They increase the energy spread and they contribute to the SR power budget of the machine. Therefore a possible strategy would be to use them is such a way that the energy spread is less than some pre-determined maximum and to switch them on only where necessary.

The maximum energy spread that can be tolerated as discussed earlier is around 52MeV. In the absence of a new design, we consider the wigglers suggested for LEP [6] that have an asymmetry of 6.15 and pole lengths of 0.65m and 4m for the strong and the weak field respectively.

The polarization time and wiggler SR power dissipated for various configurations can be seen in Table 4. In each case we have pushed the wiggler field until the maximum allowed energy spread of 52MeV is reached. B+ is the field of the strong pole. As can be seen, polarization times are reduced by a large factor - to 21 hours (TLEP) and 7 hours (CEPC) when using wigglers. Interestingly, polarization times depend only weakly on the number of wigglers installed (but a higher field per wiggler is needed)

Therefore useful polarization levels (5-10%) are reached after 60-130 minutes (TLEP) and 21-42 minutes (CEPC). These times are not too different from the fill up times of the machines.

The SR power dissipated by the wigglers (last column of Table 4, for both beams) is rather large, although it is reduced if one operates one wiggler at a high field rather than many at a reduced field.

**Table 4:** the effect of the use of wigglers on polarization times, energy spread and wiggler power dissipation using the analytic approach in **[6]** and for the wiggler design described therein. B+ is the magnetic field of the short (strong) dipole of the wiggler.

| Machine | Energy (GeV) | No. of wigglers | B+ (T) | Polarization time (hours) | Energy spread (MeV) | Wiggler SR power (MW) |
|---|---|---|---|---|---|---|
| TLEP | 45 | 0 | 0 | 253 | 17 | 0 |
| TLEP | 45 | 12 | 0.62 | 21 | 52 | 20 |
| TLEP | 45 | 1 | 1.35 | 24 | 52 | 9 |
| CEPC | 45 | 0 | 0 | 41 | 23 | 0 |
| CEPC | 45 | 12 | 0.72 | 7 | 52 | 17 |
| CEPC | 45 | 1 | 1.58 | 7 | 52 | 7 |

*Wiggler operation*

A possible strategy therefore emerges to solve the problem of very long polarization times at large storage rings while wasting as little of the power budget of the machine (which costs in terms of luminosity) as possible: Wigglers need to be used, but they need to be on just enough time to polarise enough non-colliding bunches. For the case of FCC-ee, 250 non-colliding bunches are sufficient and for the case of the CEPC 40 non-colliding bunches. The wigglers can be switched on as soon as the machine starts filling up and they can be switched off when 5% polarization is achieved. Machine fill-up times are expected to be around 30 minutes, therefore in the case of the FCC-ee an extra ~40 minute dead time is introduced while polarization builds up and during which period no meaningful energy measurement can be performed. Also, due to the power taken up by the wigglers, the luminosity of the machine will be lower than during normal operation. Physics studies which do not need the precise energy determination can take place, though.

When the required level of polarization for the non-colliding bunches has been achieved, the wigglers can be turned off and the depolarization measurements can start. Measuring and replacing 5 bunches for 5 depolarization measurements per hour, the FCC-ee will exhaust all 250 non-colliding bunches in 50 hours, during which time new non-colliding bunches will have been polarized to more than 5%.

We here assume that the number of electrons in a non-colliding bunch would be similar to the number of electrons of a normal (colliding) bunch. For the FCC-ee this number is $\sim 1.8 \cdot 10^{11}$ (similar to the LEP1 value). Having 250 out of 16700 bunches not colliding leads to an inefficiency of 1.5%.

The SR budget for the wigglers is reasonable, especially considering that they can be switched off after a short period of time. A single wiggler with a high field, if it can be constructed at a reasonable cost, is better than many wigglers with a smaller field in this respect. It should be noted here that wigglers introduce more damping and might help to achieve higher beam-beam parameters, partly compensating the luminosity loss due to wiggler SR power – this is a topic that needs to be investigated.

# CONCLUSIONS

The resonant depolarization method seems accessible at the Z (45GeV) and W (80GeV) energies of the FCC-ee and at the Z energy of the CEPC. Both lepton species should be measured. Long polarization times necessitate the use of wigglers, which however are needed only during a short period at the beginning of a fill. Resonant depolarization measurements should be performed routinely at a rate of a few her hour to profit from their instantaneous accuracy and help reduce the energy uncertainty to the unprecedented levels needed by the high statistics of the proposed large circular colliders.